\documentclass[aps,prl,twocolumn,groupedaddress,showpacs]{revtex4}

\usepackage{graphicx}

\begin{document}

\title{Atom interferometry with Bose-Einstein condensates in a double-well potential}

\author{Y. Shin}
\author{M. Saba}
\author{T. A. Pasquini}
\author{W. Ketterle}
\author{D. E. Pritchard}
\author{A. E. Leanhardt}

\homepage[URL: ]{http://cua.mit.edu/ketterle_group/}

\affiliation{Department of Physics, MIT-Harvard Center for
Ultracold Atoms, and Research Laboratory of Electronics,
Massachusetts Institute of Technology, Cambridge, Massachusetts,
02139}

\date{\today}

\begin{abstract}

A trapped-atom interferometer was demonstrated using gaseous
Bose-Einstein condensates coherently split by deforming an optical
single-well potential into a double-well potential.  The relative
phase between the two condensates was determined from the spatial
phase of the matter wave interference pattern formed upon
releasing the condensates from the separated potential wells.
Coherent phase evolution was observed for condensates held
separated by 13~$\mu$m for up to 5~ms and was controlled by
applying ac Stark shift potentials to either of the two separated
condensates.

\end{abstract}


\maketitle

Demonstrating atom interferometry with particles confined by
magnetic~\cite{OFS01,HHH01,LCK02,SKV03} and optical~\cite{DMV02}
microtraps and waveguides would realize the matter wave analog of
optical interferometry using fiber-optic devices.  Current
proposals for confined-atom interferometers rely on the merger and
separation of two potential wells to coherently divide atomic
wavepackets~\cite{HVB01,HRH01,ACF02}. This type of division
differs from previously demonstrated atomic beam splitters. To
date, atomic beams and vapors have been coherently diffracted into
different momentum states by mechanical~\cite{CAM91,KET91} and
optical~\cite{KAC91} gratings, and Bose-Einstein condensates have
been coherently delocalized over multiple sites in optical
lattices~\cite{ANK98,OTF01,CBF01,GBM01,GME02,MGW03}. Atom
interferometers utilizing these beam splitting elements have been
used to sense accelerations~\cite{PCY97,ANK98} and
rotations~\cite{LHS97,GBK97}, monitor quantum
decoherence~\cite{CHL95}, characterize atomic and molecular
properties~\cite{ESC95}, and measure fundamental
constants~\cite{PCY97,GDH02}.

In this Letter, we demonstrate a trapped-atom interferometer with
gaseous Bose-Einstein condensates confined in an optical
double-well potential. Condensates were coherently split by
deforming an initially single-well potential into two wells
separated by 13~$\mu$m.  The relative phase between the two
condensates was determined from the spatial phase of the matter
wave interference pattern formed upon releasing the atoms from the
separated potential wells~\cite{ATM97,MGW03}.  This recombination
method avoids deleterious mean field effects~\cite{SFG97,STZ02}
and detects applied phase shifts on a single realization of the
experiment, unlike in-trap recombination
schemes~\cite{HVB01,HRH01,ACF02}.

The large separation between the split potential wells allowed the
phase of each condensate to evolve independently and either
condensate to be addressed individually. An ac Stark phase shift
was applied to either condensate by temporarily turning off the
optical fields generating its potential well. The spatial phase of
the resulting matter wave interference pattern shifted linearly
with the applied phase shift and was independent of the time of
its application.  This verified the phase sensitivity of the
interferometer and the independent phase evolution of the
separated condensates.  The measured coherence time of the
separated condensates was 5~ms.

The present work demonstrates a trapped-atom interferometer with
two interfering paths.  This geometry has the flexibility to
measure either highly localized potentials or uniform potential
gradients, such as those arising from atom-surface interactions or
the earth's gravitational field, respectively. In contrast,
multiple-path interferometers demonstrated in optical lattice
systems are restricted to measurements of the
latter~\cite{ANK98,MGW03}.

Bose-Einstein condensates containing over $10^7$ $^{23}$Na atoms
were created in the $|F = 1,m_F = -1\rangle$ state in a magnetic
trap, captured in the focus of a 1064~nm optical tweezers laser
beam, and transferred into an auxiliary ``science'' chamber as
described in Ref.~\cite{GCL02}.  In the science chamber, the
condensate was loaded from the optical tweezers into the
interferometer's single-well optical trap formed by a
counter-propagating, orthogonally-polarized 1064~nm laser beam
shifted in frequency from the tweezers by $\sim 100$~MHz to avoid
interference effects.

\begin{figure}
\begin{center}
\includegraphics{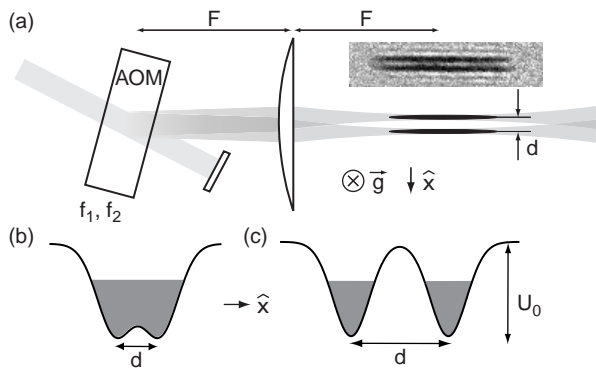}
\caption{(a) Schematic diagram of the optical setup for the
double-well potential. An acousto-optic modulator (AOM) was driven
by two frequencies, $f_{1}$ and $f_{2}$, and diffracted a
collimated beam into two beams. The AOM was placed in the focal
plane of a lens of focal length $F$ so that the two beams
propagated parallel to each other. The radial separation of the
potential wells, $d$, was controlled by the frequency difference,
$\Delta f = |f_{1}-f_{2}|$. $\vec{g}$ denotes the direction of
gravitational acceleration. The absorption image shows two
well-separated condensates confined in the double-well potential
diagramed in (c). The field of view is 70~$\mu$m $\times$
300~$\mu$m.  Energy diagrams for (b) initial single-well trap with
$d=6\ \mu$m and (c) final double-well trap with $d=13\ \mu$m.  In
both (b) and (c), $U_0 = h \times 5$~kHz and the peak atomic mean
field energy was $\sim h \times 3$~kHz. The potential ``dimple''
in (b) was $<h \times 500$~Hz which was much less than the peak
atomic mean field energy allowing the trap to be characterized as
a single-well. The potential ``barrier'' in (c) was $h \times
4.7$~kHz which was larger than the peak atomic mean field energy
allowing the resulting split condensates to be characterized as
independent.\label{f:setup}}
\end{center}
\end{figure}

A schematic diagram of the setup for the interferometer's optical
trap is shown in Fig.~\ref{f:setup}(a). The optical potentials
were derived from a collimated laser beam that passed through an
acousto-optic modulator (AOM) and was focused onto the condensate
with a lens. The AOM was driven by two radio frequency (rf)
signals to create the double-well potential. The separation
between the potential wells was controlled by the frequency
difference between the rf drives. The $1/e^2$ radius of each
focused beam was 5~$\mu$m. For typical optical powers, this
resulted in a single beam trap depth $U_0 = h \times 5$~kHz, where
$h$ is Planck's constant, and a radial (axial) trap frequency $f_r
= 615$~Hz ($f_z = 30$~Hz).

The condensate was initially loaded into the single-well trap
shown in Fig.~\ref{f:setup}(b).  After holding the cloud in this
trap for 15~s to damp excitations, the peak atomic mean field
energy was $\tilde{\mu}_0 \approx h \times 3$~kHz. The single-well
trap was deformed into the double-well potential shown in
Fig.~\ref{f:setup}(c) by linearly increasing the frequency
difference between the rf signals driving the AOM in 5~ms.  The
amplitude of the rf signals were tailored during the splitting
process to guarantee an even division of the condensate atoms and
nearly equal trap depths after splitting.

\begin{figure}
\begin{center}
\includegraphics{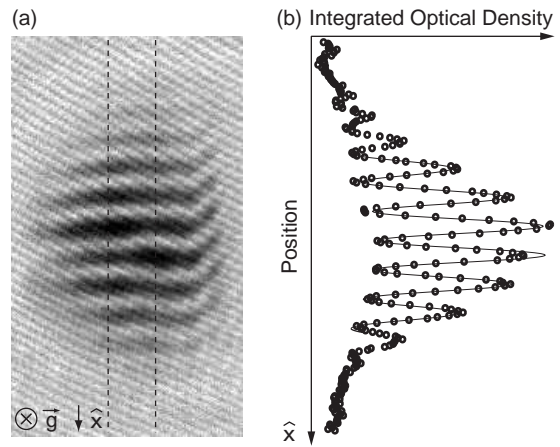}
\caption{Matter wave interference. (a) Absorption image of
condensates released from the double-well potential in
Fig.~\ref{f:setup}(c) and allowed to overlap during 30~ms of
ballistic expansion.  The imaging axis was parallel to the
direction of gravitational acceleration, $\vec{g}$.  The field of
view is 600~$\mu$m $\times$ 350~$\mu$m. (b) Radial density
profiles were obtained by integrating the absorption signal
between the dashed lines, and typical images gave $>60\%$
contrast. The solid line is a fit to a sinusoidally-modulated
Gaussian curve from which the phase of the interference pattern
was extracted (see text).\label{f:typical}}
\end{center}
\end{figure}

The key achievement of this work was the reproducibility of the
spatial phase of the matter wave interference pattern on each
realization of the experiment. Figure~\ref{f:typical} shows a
typical matter wave interference pattern formed by the condensates
released from  the double-well potential. The reproducibility
directly confirmed that deforming the optical potential from a
single-well into a double-well coherently split the condensate
into two clouds with deterministic relative phase. While past work
suffered from an unstable potential barrier separating the two
condensates and irreproducible turn off a high current magnetic
trap to initiate ballistic expansion~\cite{ATM97}, the current
experiment derived its double-well potential from a single laser
beam. Thus, vibrations and fluctuations of the laser beam were
common-mode to both wells and a clean and rapid trap turn off was
achieved.

\begin{figure}
\begin{center}
\includegraphics{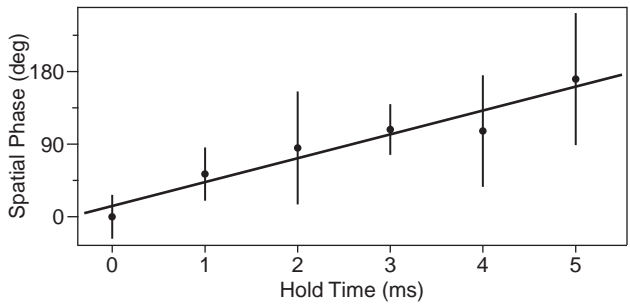}
\caption{Phase coherence of the separated condensates. The spatial
phase of the interference pattern is plotted versus hold time
after splitting. Each point represents the average of eight
measurements, and the error bars are $\pm$ one standard deviation.
The phase evolution was due to unequal trap depths for the two
wells, which was determined from the linear fit to be $h \times
70$~Hz or $\sim 1\%$ of the trap depth.\label{f:hold}}
\end{center}
\end{figure}

\begin{figure}
\begin{center}
\includegraphics{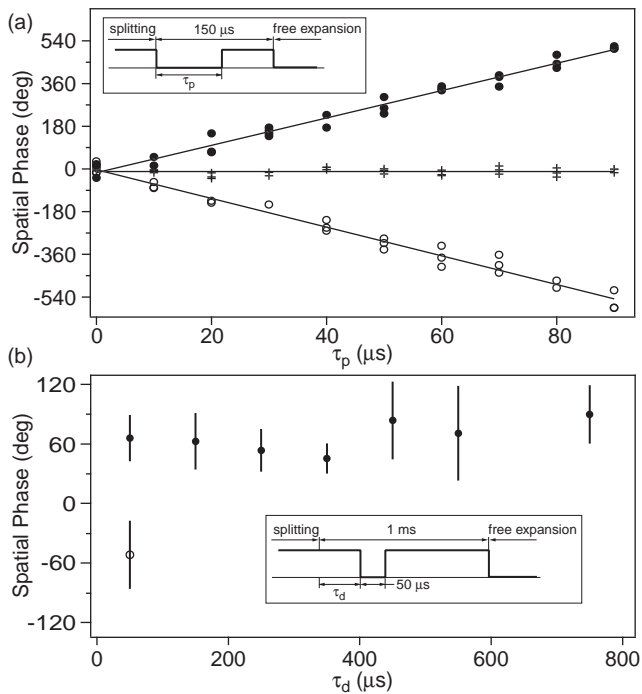}
\caption{Trapped-atom interferometry. (a) ac Stark phase shifts
were applied to either well exclusively (solid circles and open
circles) or both wells simultaneously (crosses) by turning off the
corresponding rf signal(s) driving the AOM for a duration
$\tau_p$.  The resulting spatial phase of the matter wave
interference pattern scaled linearly with $\tau_p$ and hence the
applied phase shift.  Applying the ac Stark shift to the opposite
well (solid versus open circles) resulted in an interference
pattern phase shift with opposite sign.  Applying ac Stark shifts
to both wells (crosses) resulted in no phase shift for the
interference pattern.  This data was taken with a slightly
modified experimental setup such that the trap depth of the
individual potential wells was $U_0 = h \times 17$~kHz.  (b) A
50~$\mu$s pulse induced a 70 degree shift independent of the pulse
position.  The experimental setup was as described in
Fig.~\ref{f:setup}. Solid and open circles have the same meaning
as in (a).  The insets show the time sequence of the optical
intensity for the well(s) temporarily turned off.\label{f:pulse}}
\end{center}
\end{figure}

The condensates were sufficiently separated that their phases
evolved independent of each other to the extent that no coupling
between the potential wells could be detected. This claim is
supported qualitatively by the absorption image in
Fig.~\ref{f:setup}(a) and the observation of high-contrast matter
wave interference patterns that penetrated the full atomic density
profile with uniform spatial period and no thick central
fringe~\cite{RNS97}, and quantitatively by measurements of the
phase evolution (Figs.~\ref{f:hold} and \ref{f:pulse}).

The relative phase between the two separated condensates was
determined by the spatial phase of their matter wave interference
pattern.  For a ballistic expansion time $t \gg 1/f_r$, each
condensate had a quadratic phase profile~\cite{DGP99},
$\psi_\pm(\vec{r},t)=\sqrt{n_\pm(\vec{r},t)}\exp(i\frac{m }{2\hbar
t}|\vec{r}\pm\vec{d}/2|^2+\phi_\pm)$, where $\pm$ denotes one well
or the other, $n_\pm$ is the condensate density, $\hbar = h /
2\pi$, $m$ is the atomic mass, and $\phi_\pm$ is the condensate
phase.  This resulted in a total density profile for the matter
wave interference pattern
\begin{equation}
\label{e:densityprofile}
    n(\vec{r},t)=(n_+ + n_- + 2\sqrt{n_+ n_-}\cos(\frac{m d}{\hbar t}x + \phi_r)),
\end{equation}
where $\phi_r=\phi_+ - \phi_-$ is the relative phase between the
two condensates and $\vec{d}=d\hat{x}$. To extract $\phi_r$, the
integrated cross section shown in Fig.~\ref{f:typical}(b) was fit
with a sinusoidally-modulated Gaussian curve, $G(x)=A
\exp(-(x-x_c)^2/\sigma^2)(1+B\cos(\frac{2\pi}{\lambda}(x-x_0) +
\phi_{f}))$, where $\phi_f$ is the phase of the interference
pattern with respect to a chosen fixed $x_0$. Ideally, if $x_0$
was set at the center of the two wells, then $\phi_r =\phi_f$.
However, misalignment of the imaging axis with the direction of
gravitational acceleration created a constant offset,
$\phi_f=\phi_r+\delta\phi$. With $t=30$~ms the measured fringe
period, $\lambda=41.5\ \mu$m, was within $4\%$ of the point source
formula prediction [Eq.~(\ref{e:densityprofile})], $h t/m d =
39.8\ \mu$m.

The coherent phase evolution of the split condensates is displayed
quantitatively in Figs.~\ref{f:hold} and \ref{f:pulse}. The
relative phase, $\phi_r$, between the separated condensates was
observed to evolve linearly in time and the standard deviation of
eight measurements of $\phi_f$ was $< 90$ degrees up to 5~ms after
splitting (Fig.~\ref{f:hold}). Furthermore, for hold times between
0 and 1~ms, the standard deviation was substantially smaller, $<
40$ degrees. Since $\phi_r$ distributed randomly between $-180$
and $+180$ degrees would have a standard deviation of $\sim 104$
degrees, the results in Figs.~\ref{f:hold} and \ref{f:pulse}
clearly demonstrate that the separated condensates had a
reproducible relative phase after splitting. The linear time
evolution of $\phi_r$ was due to a chemical potential difference
between separated condensates and could be controlled by varying
the trap depths of the individual potential wells after splitting.

Fundamental limits on the phase coherence between isolated
condensates arise due to Poissonian number fluctuations associated
with the coherent state description of the
condensate~\cite{LEY96,JAW97,MAC01}.  For our experimental
parameters, the time scale for phase diffusion was $\sim 200$~ms.
The uncertainty in determining $\phi_f$ at longer hold times
$>5~$ms is attributed to axial and breathing-mode excitations
created during the splitting process.  These excitations lead to
interference fringes that were angled and had substantial
curvature, rendering a determination of $\phi_f$ impossible.
Splitting the condensate more slowly in an effort to minimize
excitations, but still fast compared to the phase diffusion time,
did not improve the measured stability of $\phi_f$. Since
controlling axial excitations appears critical for maintaining
phase coherence, splitting condensates that are freely propagating
in a waveguide potential may be more promising~\cite{LCK02}.

The phase sensitivity of the trapped-atom interferometer was
demonstrated by applying ac Stark phase shifts to either (or both)
of the two separated condensates.  Phase shifts were applied to
individual condensates by pulsing off the optical power generating
the corresponding potential well for a duration $\tau_p \ll
1/f_r$. Figure 4(a) shows that the spatial phase of the matter
wave interference pattern shifted linearly with the pulse
duration, as expected. Due to the inhomogeneous optical potential,
$U(r)$, the applied ac Stark phase shifts varied across the
condensate as $\Delta\phi(r) = -U(r)\tau_p/\hbar$. Averaging this
phase shift over the inhomogeneous condensate density,
$n(\vec{r})$, approximates the expected spatial phase shift of the
matter wave interference pattern as $\Delta
\bar{\phi}=\frac{1}{N}\int
d^3\vec{r}~n(\vec{r})\Delta\phi(\vec{r}) =
(U_0-\frac{2}{7}\tilde{\mu}_0)\Delta t /\hbar$, where $N$ is the
number of atoms, and $U_0$ and $\tilde{\mu}_0$ are the trap depth
and mean field energy at the center of each potential well,
respectively.  The measured phase shifts yielded $U_0=h\times
5$~kHz [Fig.~\ref{f:pulse}(b)], which was consistent with
calculations based on other measured trap parameters.

The measured phase shifts of the matter wave interference depended
only on the time-integral of the applied ac Stark phase shifts
[Fig.~\ref{f:pulse}(b)], as expected for uncoupled condensates.
The final relative phase, $\phi_r$, should be the same on
different phase trajectories because the history of phase
accumulation does not affect the total amount of accumulated
phase. For coupled condensates, Josephson oscillations between the
wells would cause the relative phase to vary nonlinearly with
time~\cite{SFG97,DGP99} and produce a time dependent signal in
Fig.~\ref{f:pulse}(b). Due to the large well separation and mean
field energy $h \times 1.7$~kHz below the barrier height, the
single-particle tunnelling rate in our system was extremely low
($\lesssim 10^{-3}$~Hz)~\cite{DGP99}, and the condensates were
effectively uncoupled.

In conclusion, we have performed atom interferometry with
Bose-Einstein condensates confined in an optical double-well
potential.  A coherent condensate beam splitter was demonstrated
by deforming a single-well potential into a double-well potential.
The large spatial separation between the potential wells
guaranteed that each condensate evolved independently and allowed
for addressing each condensate individually. Recombination was
performed by releasing the atoms from the double-well potential
and allowing them to overlap while expanding ballistically.
Implementing a similar readout scheme with magnetic potentials
generated by microfabricated current carrying wires should be
possible and would eliminate deleterious mean field effects
inherent in proposals using in-trap wavepacket recombination.
Propagating the separated condensates along a waveguide prior to
phase readout would create an atom interferometer with an enclosed
area, and hence with rotation sensitivity.

We thank W.\ Jhe, C.\ V.\ Nielsen, and A.\ Schirotzek for
experimental assistance and S.\ Gupta, Z.\ Hadzibabic, and M.\ W.\
Zwierlein for critical comments on the manuscript. This work was
funded by ARO, NSF, ONR, and NASA.  M.S.\ acknowledges additional
support from the Swiss National Science Foundation.

\end{document}